\documentclass[10pt,journal,compsoc]{IEEEtran}
\usepackage[tablegrid]{vhistory}
\usepackage{microtype}
\usepackage{mathptmx}
\usepackage{amsmath,amsfonts,bm}
\usepackage[english]{babel} 
\usepackage{listings}
\usepackage{hyperref}
\usepackage[nocompress]{cite}
\usepackage{graphicx}

\usepackage[listings,skins,raster]{tcolorbox}

\usepackage{astrojournals}

\newcommand\pytorch{%
  {\tt PyTorch}}

\definecolor{color1}{RGB}{244,243,224}
\definecolor{color2}{RGB}{31,174,174}
\newtcolorbox{boxnumpy}[1][]{%
  colback=yellow!5,
  colframe=color2,
  colbacktitle=color1,
  coltitle=black,
  title={\parbox[t]{2cm}{NumPy:}\parbox[t]{\dimexpr\linewidth-2cm\relax}{#1}},
  fonttitle=\ttfamily,
  width=0.5\linewidth,
  boxsep=1pt,
  left=1pt
}

\newtcolorbox{boxtorch}[1][]{%
  colback=blue!5,
  colframe=color2,
  colbacktitle=blue!7,
  coltitle=black,
  title={\parbox[t]{2cm}{\pytorch:}\parbox[t]{\dimexpr\linewidth-2cm\relax}{#1}},
  fonttitle=\ttfamily,
  width=0.5\linewidth,
  boxsep=1pt,
  left=1pt  
}

\renewcommand{\vec}[1]{\ensuremath{\bm{#1}}}

\title{Acceleration of Non-Linear Minimisation with {\emph{PyTorch}}}

\author{Bojan Nikolic\\Astrophysics Group, Cavendish
  Laboratory, University of Cambridge, UK\thanks{\url{email:b.nikolic@mrao.cam.ac.uk}}}

\IEEEtitleabstractindextext{%
\begin{abstract}
  I show that a software framework intended primarily for training of
  neural networks, PyTorch, is easily applied to a general function
  minimisation problem in science.  The qualities of PyTorch of
  ease-of-use and very high efficiency are found to be applicable in
  this domain and lead to two orders of magnitude improvement in
  time-to-solution with very small software engineering effort. 
\end{abstract}
}

\begin{document}
\maketitle

\IEEEraisesectionheading{\section{Introduction}\label{sec:introduction}}

Minimisation (or, equivalently, maximisation) of non-linear functions
is a widespread tool in science, engineering and information
technology. Common examples of its use are maximum likelihood and
maximum a-posteriori estimates of model parameters given a set of
measurements; and, optimisation of system designs by minimising a cost
function.

The unconstrained minimisation problem is posed as:
\begin{equation}
  \operatorname*{argmin}_{\vec{x} \in R^{n} } \,\, f(\vec{x})
\end{equation}
where $f$ is the single-valued function to be minimised and its $n$
parameters are expressed as if they were components of vector
$\vec{x}$.  Most non-linear minimisation algorithms require the
gradient of the function to be minimised \cite{Nocedal2006NO}, i.e.:
\begin{equation}
  \nabla  f = \frac{\partial f}{\partial x_i}.
\end{equation}

In a wide class of problems it is expensive (in terms of computational
resources) to compute $f(\vec{x})$ and $ \nabla f$ meaning the
minimisation as a whole is expensive. Additionally, there are
time-sensitive applications, e.g., in control systems, where the
latency between observation and the solution is important even if the
overall expense is not. For these reasons it is desirable to
accelerate minimisation as a whole and in particular the evaluation of
$f(\vec{x})$ and $ \nabla f$. At the same time, the function to be
minimised and its gradient can be very complex (and subject to
evolution over time), yet it is \emph{essential} that they are
implemented correctly. For this reasons, extensive by-hand
optimisation of the code of their implementation is undesirable as
this is typically inflexible and error prone.

Training of neural networks can also be expressed as a minimisation
problem. Although the algorithms used for this minimisation are
sometimes specific (to take into account that only a part of the
available training set is considered at a time), they usually involve
frequent evaluation of potentially expensive $f(\vec{x})$ and
$ \nabla f$. Rapid adoption of neural networks in information
technology systems has lead to significant investment into software to
support their training, including \pytorch. The new software packages
developed in this area emphasise both efficiency and ease of use.

In this note I show \pytorch\ can easily be used for general
minimisation problems in science and that its qualities of ease of use
and efficiency are maintained.

\section{Why \pytorch ? }

\pytorch\footnote{\url{https://pytorch.org/}} has four key features
which make it particularly suitable for accelerating non-linear
function minimisation in science and engineering:
\begin{enumerate}
\item The interface is in Python, modelled after the widely used NumPy
  \cite{Oliphant20074160250} library, and introduces very few
  restrictions on the programmer
\item It supports automatic reverse-mode differentiation for very
  efficient computation of $ \nabla f$ without user intervention
\item Easy-to-use offloading of operations onto Graphical Processing
  Units (GPUs) for efficient computation of data-parallel operations
\item It is available as high-quality open-source distribution
\end{enumerate}
These features combine into a programming environment that is familiar
to engineers and scientists while enabling accurate and high
performance minimisation. At the time this study was begun \pytorch\
had a clear advantage in these features versus the popular {\tt
  TensorFlow} system; however {\tt TensorFlow} has recently introduced
`eager' mode which supports dynamic, imperative computational graphs
and automatic differentiation, and which therefore should work close
to equivalently to \pytorch.

\subsection{Automatic differentiation}

Automatic differentiation is distinct from symbolic and numerical
differentiation. It is based on applying the chain rule to accumulate
the numerical value of the differential as the algorithm executes (or,
in the case of reverse-mode, tracing back along the algorithm
execution path). Its key advantages in the present case are that:
\begin{enumerate}
\item The computational cost of evaluating $ \nabla f$ is (in the
  reverse mode) of order the cost evaluating $f(\vec{x})$ independent
  of the number of parameters of the function. This is in contrast to
  numerical differentiation, the cost of which scales as $n$ where $n$
  is the number of parameters. 
\item It requires no implementation from the application programmer
\item Support of new elemental functions is relatively easy as no
  symbolic algebra manipulation is required
\end{enumerate}
Use of automatic differentiation for obtaining function gradients in
function minimisation is established\cite{Nocedal2006NO}, but
\pytorch\cite{paszke2017automatic} combines it with ease-of-use and
automatic offloading onto GPUs. A review of the application of
automatic differentiation in training of neural networks is given by
\cite{2015arXiv150205767G}.

\subsection{Interface}

\pytorch\ presents an interface based around multi-dimensional arrays
(`tensors') and defines operations over whole arrays or their
subsets. The interface is similar to NumPy. The key feature of this
interface is that enables the application programmer to express the
algorithm in terms of high-level operations and avoiding explicit,
serial, iteration. This in turn enables acceleration and automatic
differentiation, while allowing the programs to be expressed in a
concise and readable form.

\subsection{GPU offloading}

\pytorch\ functions and operations have implementations defined for
NVIDIA/CUDA GPUs as well as conventional CPUs. The application level
programmers responsibility is to trigger movement of objects between
GPU and main memories; all operations on these objects are done by
implementations appropriate for their current location.  This design
provides for easy offloading of computations onto GPUs without
opportunity for introduction of errors.  GPUs have a substantially
higher execution throughput for a wide range of data parallel
computations.

\section{Example application: Maximum Likelihood Phase Retrieval}

\begin{figure*}
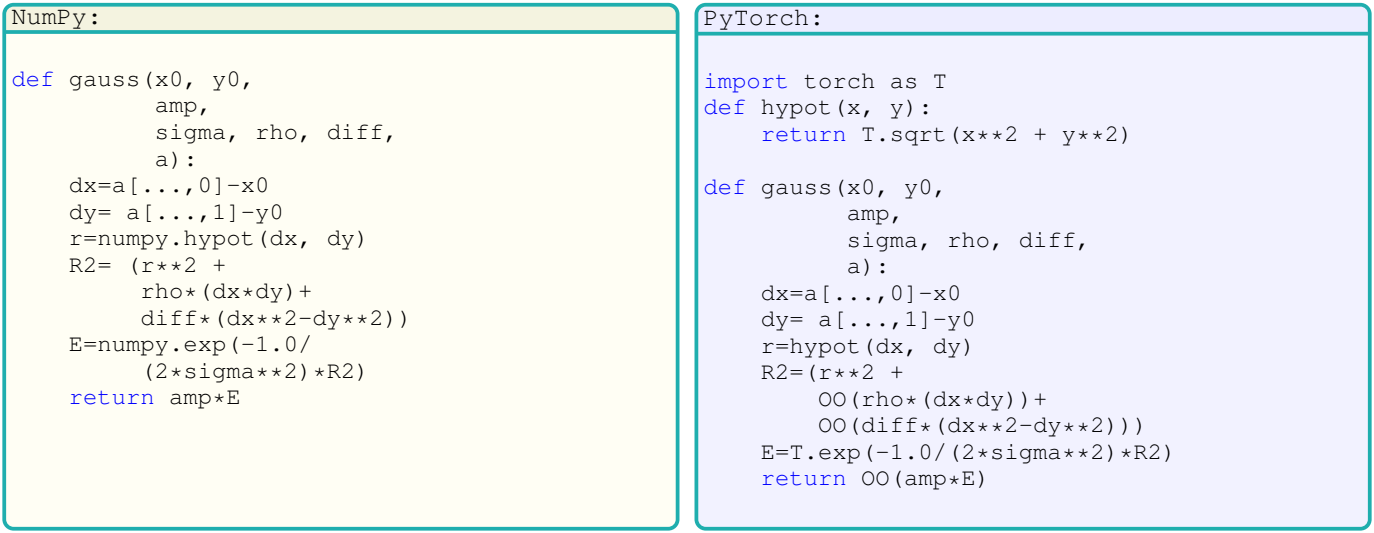

\begin{tcbraster}[raster columns=2,raster equal height]
  \begin{boxnumpy}[]
      \begin{lstlisting}
def gauss(x0, y0,
          amp,
          sigma, rho, diff,
          a):
    dx=a[...,0]-x0
    dy= a[...,1]-y0
    r=numpy.hypot(dx, dy)
    R2= (r**2 +
         rho*(dx*dy)+
         diff*(dx**2-dy**2))
    E=numpy.exp(-1.0/
         (2*sigma**2)*R2)
    return amp*E
  \end{lstlisting}
\end{boxnumpy}
\begin{boxtorch}[]
  \begin{lstlisting}
import torch as T    
def hypot(x, y):
    return T.sqrt(x**2 + y**2)

def gauss(x0, y0,
          amp,
          sigma, rho, diff,
          a):
    dx=a[...,0]-x0
    dy= a[...,1]-y0
    r=hypot(dx, dy)
    R2=(r**2 +
        OO(rho*(dx*dy))+
        OO(diff*(dx**2-dy**2)))
    E=T.exp(-1.0/(2*sigma**2)*R2)
    return OO(amp*E)

\end{lstlisting}
\end{boxtorch}
\end{tcbraster}
\caption{Comparison of NumPy and \pytorch\ implementations of a
  function that models the Gaussian illumination (or apodisation) of
  the aperture plane. The input parameters to the Gaussian function
  are it position (x0, y0), amplitude (amp), shape (sigma, rho and
  diff), and the raster on which it is to be evaluated (a). Function
  \lstinline|OO| handles to optional offload onto GPUs. }
\label{fig:compgaus}
\end{figure*}

I illustrate the application of \pytorch\ to function minimisation in
science by applying it to the phase retrieval problem. Phase retrieval
is the derivation of the phase of an oscillatory field
(electromagnetic or particle-wave) from far-field measurements of its
power only. It is a common technique used in demanding imaging
scientific applications.  I here show an application to a
maximum-likelihood, model-based, phase retrieval in radio astronomy
used to measure and optimise some of the largest single-dish radio
telescopes\cite{2007A&A...465..679N}. This specific technique is
called Out-Of-Focus (OOF) holography: out-of-focus because the optical
system is intentionally defocused to introduce a known phase change
(`phase diversity') into it; holography because, as in Gabor's sense,
both the amplitude and the phase of the field are ultimately obtained.

\subsection{Relation between maximum-likelihood and function minimisation}

Likelihood is the probability $P(Y | \vec{x}, M)$ of observing a
particular measurement set $Y\equiv \{{y_i\}}$ given a model M and its
parameters $\vec{x}$. The maximum likelihood approach estimates the
true value of model parameters $\vec{\hat{x}}$ as those that maximise
the likelihood of observing the data:
\begin{equation}
  \operatorname*{argmax}_{\vec{x}} \,\, P(Y | \vec{x}, M).
\end{equation}
The maximisation (or equivalently, minimisation of the negative
log-likelihood), can in general be a global problem, but here I
consider the local version in which the maximum in a neighbourhood is
sought.

Maximum a-posteriori (MAP) estimation of model parameters augments the
maximum likelihood approach but also considering the prior probability
distributions of model parameters. Such estimators are also used in
image reconstruction, e.g., \cite{Sarder2006} and
\cite{ao4elt3_13290}. The technique presented here is also directly
applicable to the MAP estimation.

A significant simplification can be made if each measurement
independent deviates from the model value according to a normal
distribution, i.e.,:
\begin{equation}
  P(y_i | \hat{y_i} ) = \frac{1}{\sqrt{2\pi \sigma^2}}
  e^{\frac{\left(y_i-  \hat{y_i} \right)^2}{2 \sigma^2}}
  \end{equation}
  where $y_i$ is the $i$-th observation, $\hat{y_i}$ is the model
  value and $\sigma$ is the measurement noise. In this case the
  minimisation of negative log-likelihood is equivalent to a
  least-squares problem which can be efficiently by the
  Levenberg-Marquardt algorithm (see for example
  \cite{Nocedal2006NO}).

  While I expect \pytorch\ will provide some acceleration to the
  non-linear least-squares Levenberg-Marquardt solver here I only
  consider the more general maximum likelihood problem to which
  Levenberg-Marquardt is not applicable.

\subsection{Maximum likelihood phase retrieval}

In the OOF holography technique the field phase at the aperture of the
telescope is parameterised as a linear combination of Zernike
polynomials. A model is established which transforms this aperture
plane field into a prediction of measurements of the far-field
response of the telescope system. The phase retrieval is done by
finding the maximum likelihood linear combination of Zernike
polynomials given a set of measurements of the far-field patter at
different focus settings. In contrast to model-free phase retrieval
techniques this technique has the advantage of good accuracy with even
relatively low dynamic range input measurement.

In the original paper \cite{2007A&A...465..679N} normally distributed
independent measurement error was assumed leading to a least-squares
formulation. Here we consider the more complex extension to a Cauchy
likelihood function:
\begin{equation}
  P(y_i | \hat{y_i} ) = \frac{1}{\pi \gamma}
    \frac{\gamma^2}{\left(y_i-  \hat{y_i}\right)^2 + \gamma^2}
  \end{equation}
  where $\gamma$ is represents the noise of measurements. This
  distribution is physically motivated for the radio astronomy
  phase-retrieval because in the very high dynamic range regime in
  which the input data are observed, errors in pointing or incaccurate
  atmospheric substraction can lead to non-Gaussian errors with wide
  tails. Similar consideration apply in a range of maximum-likelihood
  problems where there is a possibility for a few measurements with
  very high error, due to e.g., glitches in read out systems, external
  unpredictable events (e.g., cosmic ray hits), etc. 

  Each input dataset consists of a number of measurements,
  which we assume have independent and uncorrelated errors; the final
  likelihood is then:
\begin{equation}
  P(\{y_{i}\}) = \prod_i  \frac{1}{\pi \gamma}
    \frac{\gamma^2}{\left(y_i-  \hat{y_i}\right)^2 + \gamma^2} 
  \end{equation}
  or more conveniently for numerical calculation
\begin{equation}
    \log P(\{y_{i}\}) = \sum_i  \log\left[ \frac{1}{\pi \gamma}
    \frac{\gamma^2}{\left(y_i-  \hat{y_i}\right)^2 + \gamma^2} \right].
\end{equation}
In the maximum likelihood approach we assume the model we use $M$ is
the correct model therefore:
\begin{equation}
  \{\hat{y}_{i}\}= M(\vec{x})
\end{equation}
where $\vec{x}$ are the model parameters as before. The computational
problem is finding the maximum of $\log L$ with respect to the
parameters $\vec{x}$. 

\subsection{Implementation}

The model M(\vec{x}) consists, briefly, of calculating a rasterised
aperture-plane field distribution which is then transformed to the
far-field power pattern using an FFT and squaring the result. The
aperture-plane field amplitude is modelled as a general
two-dimensional Gaussian function, while the aperture-plane phase is
modelled as a linear combination of Zernike polynomials with an
additional defocus term when modelling out-of-focus beam measurements.
Implemented in NumPy, the model is of moderate complexity: around 100
lines of code excluding comments and blanks (an earlier C++
implementation required an order of magnitude many more lines of
code).

Transforming the model into a \pytorch\ implementation took around
four hours of programming time and produced a model of comparable
complexity and readability. Several reasons were observed for the
relatively low amount of work: firstly, only parts of the models that
are functions of the model parameters need to be translated in
\pytorch, meaning some relatively complex parts such as calculating
the rasterised Zernike polynomials did not need translation at
all. Secondly, most NumPy functions have direct counterparts in
\pytorch; those did not have counterparts were all written in NumPy
themselves (not in a C extension) and could easily be re-implemented
in \pytorch\ with reference to the their source code.

A typical translation into \pytorch\ is illustrated in
Figure~\ref{fig:compgaus} with the implementation of the Gaussian
amplitude model. The \pytorch\ version is essentially the same the
NumPy version except that:
\begin{enumerate}
\item The \lstinline|numpy.hypot| does not-have a built-in equivalent,
  so is (trivially) re-implemented in \pytorch\ 
\item Using the \lstinline|OO| custom function to optionally offload
  data structures onto GPUs. This function uses a global variable to
  decide if the \lstinline|.cuda()| method of a \pytorch\ tensor will
  be called.
\end{enumerate}

\begin{figure*}
\begin{tcbraster}[raster columns=2,raster equal height]
  \begin{boxnumpy}[]
      \begin{lstlisting}
def tosky(a,p):
    as= a* numpy.exp(1j*p)
    s=numpy.fft.ifft2(as)
    return s.real**2 + s.imag**2
  \end{lstlisting}
\end{boxnumpy}
\begin{boxtorch}[]
  \begin{lstlisting}
import torch as T    

def abs2(x):
    return (x**2).sum(dim=-1)

def tosky(a, p):
    c=a*T.cos(p)
    s=a*T.sin(p)
    S=T.ifft(T.stack([c,s],
                     dim=-1),
                 signal_ndim=2,
                 normalized=False)
    return abs2(S)

\end{lstlisting}
\end{boxtorch}
\end{tcbraster}
\caption{Comparison of NumPy and \pytorch\ implementations of a function
  to transform the aperture plane field to far-field power diffraction
  pattern. The input parameters are \lstinline|a|, amplitude of the
  aperture-plane field, and \lstinline|p|, phase of the field.}
\label{fig:comptosky}
\end{figure*}

A more complex translation is shown in Figure~\ref{fig:comptosky} with
the code to transform an aperture plane field distribution to the
far-field.  The reason for the additional complexity is \pytorch\ does
not support complex numbers and so in the translation the real and
imaginary parts need to be represented as an additional, innermost,
dimension of the arrays.

The same likelihood function can be used for both NumPy and \pytorch\ 
versions as it uses only basic mathematical operations (addition,
division and raising to a power) which are overloaded for both NumPy
and \pytorch\ arrays.

Maximisation of the likelihood is done using the
Broyden-Fletcher-Goldfarb-Shanno (BFGS) algorithm in both
variants. This is a local algorithm, i.e., it does not attempt to find
global function maxima. This is consistent with the original
implementation by \cite{2007A&A...465..679N} and is supported by
empirical observations over the intervening years that local
minimisation converges to the correct solution for this problem.  The
implementation used is the SciPy optimisation function
\lstinline|scipy.optimize.minimize| with \lstinline|method=BFGS|. The
only difference between the NumPy and \pytorch\ is that with \pytorch\
we have available cheaply and automatically evaluated Jacobian of the
function, so the optimisation is done with the \lstinline|jac=True|
flag.

The complete source code is publicly available at
\url{https://github.com/bnikolic/oof/tree/ooftorch} in the
``ooftorch'' branch.

\subsection{Performance Measurement}

Relative performance of the NumPy and \pytorch\ variants were measured
on a dedicated Dell PowerEdge server with dual socket Intel Xeon
E5-2630 CPUs and dual NVIDIA Tesla K20c GPUs.  Measurement was made as
function of:
\begin{enumerate}
\item Number of model parameters, represented by the maximum order of
  Zernike polynomials used, $n$. The actual number of polynomials up
  to order $n$ is $n(n+3)/2 +1$, so for example if Zernike polynomials
  of to order $n=8$ are used there are 45 parameters to be optimised.
\item Computational cost of the model calculation, represented number
  of pixels $N$ in each dimension of the grid.
\end{enumerate}
Simulated measurements (including simulated noise) were used as an
input into the phase retrieval and it was found that the different
implementations converged to the same result up to the tolerances
specified to the BFGS algorithm.

\begin{figure}
  \includegraphics[width=\columnwidth]{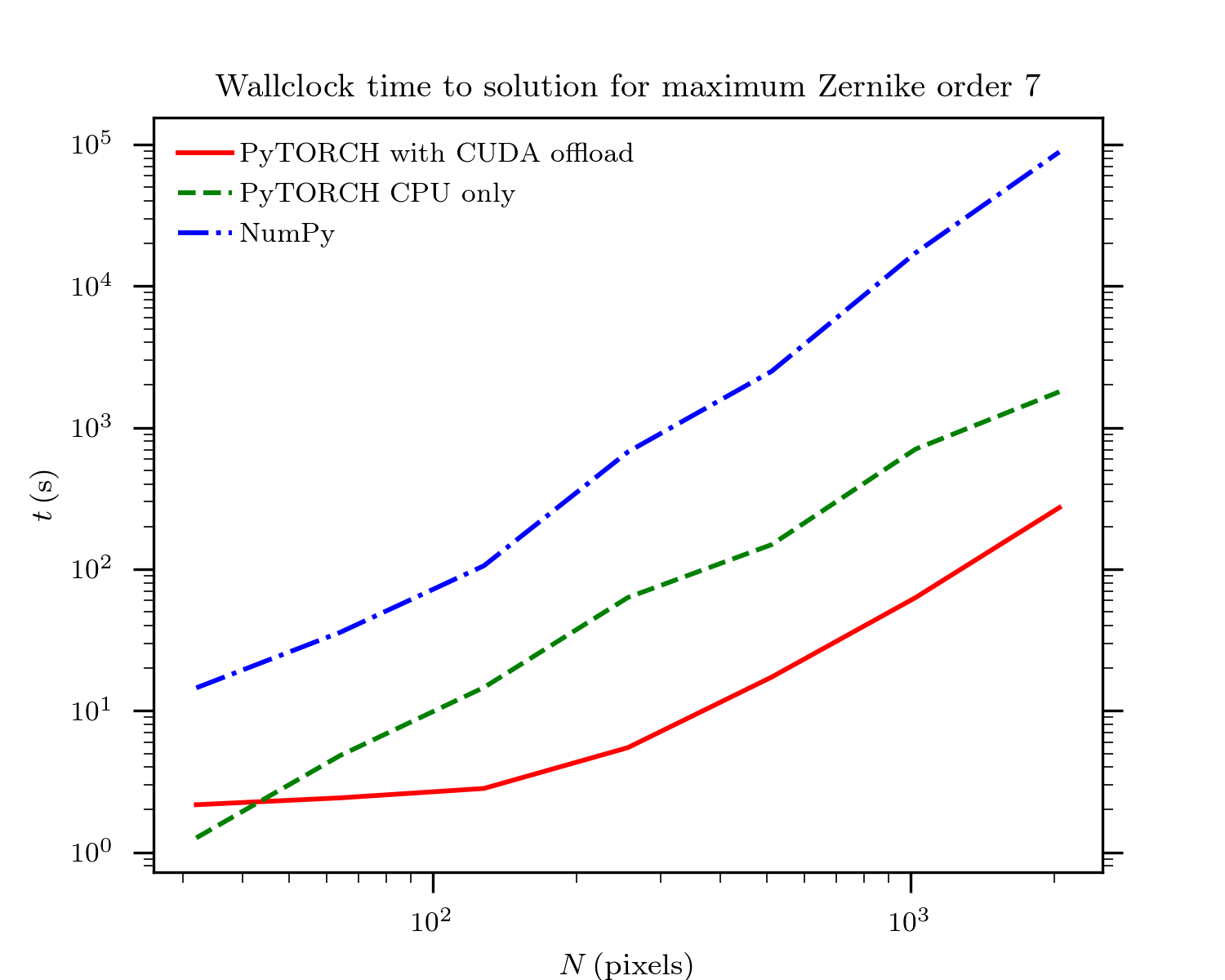}
  \caption{Time to solution as function of the grid size used for
    modelling the telescope optics. All measurements on a dual-socket
    Intel Xeon CPU E5-2630 with dual NVidia Tesla K20c GPUs.  {\bf
      NumPy}: plain CPU-only NumPy implementation; {\bf PyTorch CPU
      only}: implementation using \pytorch\ but without using offloading
    onto the GPU; {\bf PyTorch with CUDA offload}: implementation
    using \pytorch\ and selecting offloading onto GPUs using the
    \pytorch\ CUDA module.}
  \label{fig:perfgrid}
\end{figure}

Measured time-to-solution as a function of grid size is shown in
Figure~\ref{fig:perfgrid}. It can be seen that the \pytorch\
implementation is far faster in all configuration, and that the
GPU-offloaded execution is faster than the CPU-only execution above
grid size $N\geq 64$. For intermediate and large grids ($N>256$) the
\pytorch\ implementation running on CPUs is approximately an order of
magnitude faster than the NumPy implementation, while the
GPU-offloaded execution is an order of magnitude faster still.

\begin{figure}
  \includegraphics[width=\columnwidth]{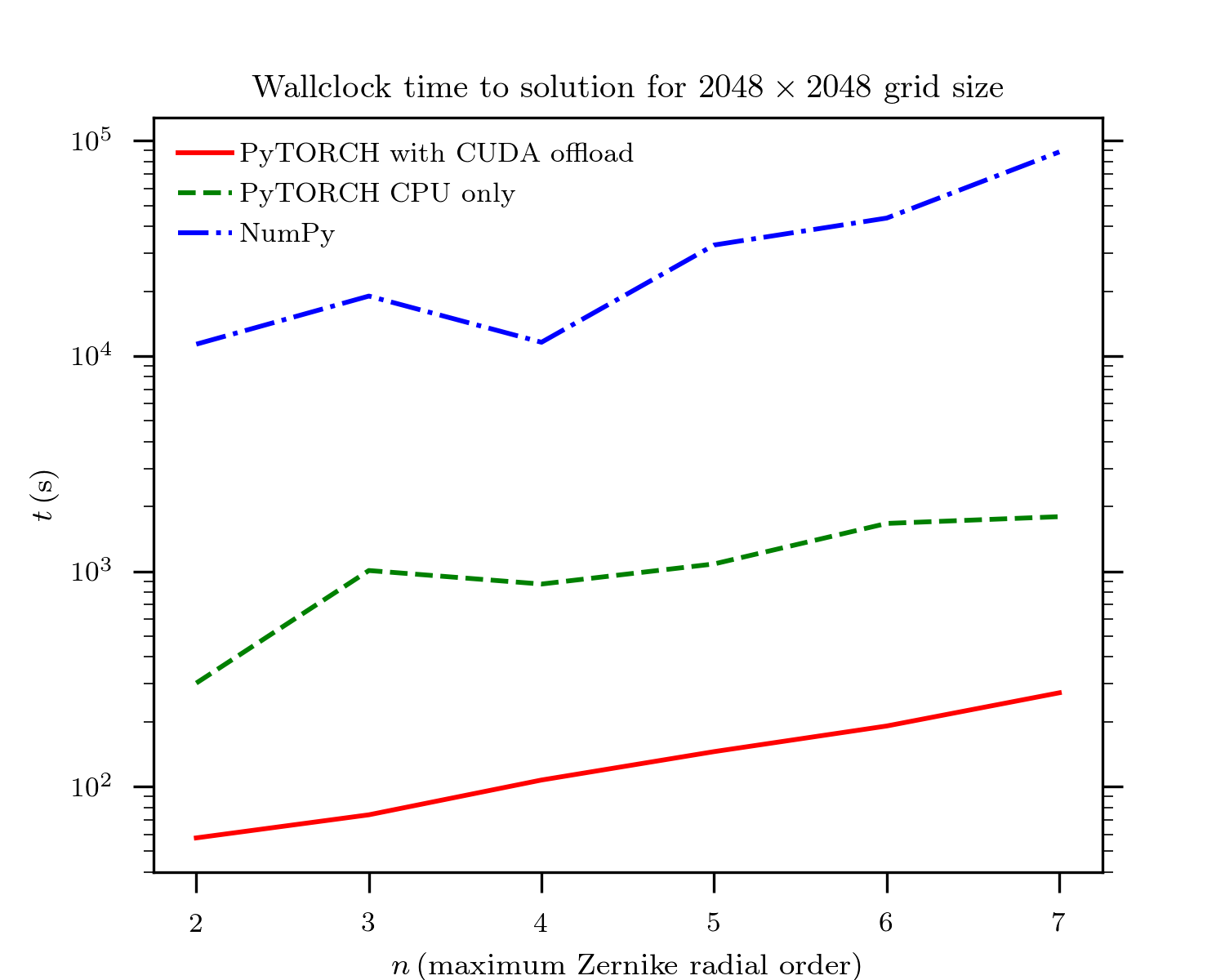}
  \caption{Time to solution as function of the maximum Zernike
    polynomial order. Measurements were performed as described in
    Figure~\ref{fig:perfgrid}.}
  \label{fig:perfnzern}
\end{figure}

Measured time-to-solution as a function of the number of model
parameters is shown in Figure~\ref{fig:perfnzern}. The large
improvement in speed by \pytorch\ is seen to be apparently independent
of the model parameters at least in this case.

\section{Related work}

All of the techniques presented here are well established (although
automatic differentiation mostly in specialised applications). The
contribution of this paper is illustrate that machine-learning
software packages combine them into an easy-to-use system that is also
highly applicable for general function minimisation.

The closest related work that I'm aware of is
\cite{2018arXiv180107232R}, where it is shown that a recurrent neural
network (RNN) model is general enough to model the finite difference
wave equation propagation in seismic modelling, and therefore that the
RNN training software ({\tt TensorFlow} in this case) can be used to
solve such problems (including use of automatic differentiation). In
contrast, in this paper I consider more general optimisation and avoid
use of a neural network model. Instead basic array operations only are
used. Furthermore in this paper focus here is on ease of use and
computational efficiency and hence the results for those are
presented. Another specialised use of \pytorch\ for fitting is
presented by \cite{BelbutePeres2017}.

NumPy as an easy-to-use and relatively efficient idiom for scientific
computing is well documented, e.g., \cite{5725236}.  There a number of
automatic differentiation tools available in Python which are studied
in relation to function optimisation by \cite{2016arXiv160606311T}.

\section{Conclusions}

I present in this paper some of the advantages that applying \pytorch\
to general function minimisation in science can bring. These
advantages are demonstrated on a realistic scientific problem,
maximum-likelihood phase retrieval, where it is shown that two orders
of magnitude of improvement in the time-to-solution can be achieved
with minimal software engineering effort. Such an improvement in
time-to-solution may open new application areas such as optimisation
of designs with many more parameters then would by typically used
today, or inclusion of fitting of complex models into low-latency
control loops.

Importantly, \pytorch\ programs are about as readable and maintainable
as NumPy programs. This is critical many minimisation and fitting
applications as subtle error in the model are often masked by
counteracting error in the best-fitting parameters. 

\section*{Acknowledgements}

I am pleased to acknowledge the support of the EC ASTERICS Project
(Grant Agreement no. 653477).

\addcontentsline{toc}{section}{References}

\bibliographystyle{IEEEtran}

\bibliography{pytorch}

\end{document}